\documentclass[transmag]{IEEEtran}
\usepackage{amsthm}
 \usepackage{amsmath}
 \usepackage{amssymb}
\usepackage{mathrsfs}
 \usepackage{mathtools}

\usepackage{cite}

\ifCLASSINFOpdf
   \usepackage[pdftex]{graphicx}
  \DeclareGraphicsExtensions{.pdf,.jpeg,.png}
\else
   \usepackage{graphicx}
   \DeclareGraphicsExtensions{.eps}
\fi

\usepackage{amsmath}
\usepackage{algorithmic}

\usepackage{array}

\ifCLASSOPTIONcompsoc
  \usepackage[caption=false,font=normalsize,labelfont=sf,textfont=sf]{subfig}
\else
\usepackage[caption=false,font=footnotesize]{subfig}
\fi
\usepackage{fixltx2e}

\usepackage{stfloats}
\usepackage{xcolor}

\newcommand{\MIMOFull}{{\tt{MIMO-FBE}}}
\newcommand{\MIMOPartial}{{\tt{MIMO-PBE}}}
\newcommand{\comment}[1]{ }

{\color{red}}

\begin{document}

\title{Deep Learning-Enabled Zero-Touch Device Identification: Mitigating the Impact of Channel Variability Through MIMO Diversity}

\author{Bechir Hamdaoui${^\dag}$, Nora Basha${^\dag}$, Kathiravetpillai Sivanesan${^\ddag}$~\\
 $^\dag$ \small Oregon State University, Corvallis, OR, USA; Email: \{hamdaoui,bashano\}@oregonstate.edu \\
  ${^\ddag}$ \small Intel Corporation, Hillsboro, OR, USA; Email: kathiravetpillai.sivanesan@intel.com 
}

\maketitle

\begin{abstract}
Deep learning-enabled device fingerprinting has proven efficient in enabling automated identification and authentication of transmitting devices. It does so by leveraging the transmitters' unique features that are inherent to hardware impairments caused during manufacturing to extract device-specific signatures that can be exploited to uniquely distinguish and separate between (identical) devices. 
Though shown to achieve promising performances, hardware fingerprinting approaches are known to suffer greatly when the training data and the testing data are generated under different channels conditions that often change when time and/or location changes. To the best of our knowledge, this work is the first to use MIMO diversity to mitigate the impact of channel variability and provide a channel-resilient device identification over flat fading channels. Specifically, we show that MIMO can increase the device classification accuracy by up to about $50\%$ when model training and testing are done over the same channel and by up to about $70\%$ when training and testing are done over different fading channels. 

\end{abstract}


\section{Introduction}
As the number of wireless devices and networks connected to the Internet keeps increasing at unprecedented scales, then so does the attack surface of such networks~\cite{lai2022healthcare}. 
Therefore, the need for automated and zero-touch approaches that can identify and authenticate devices based on signatures that are immune to spoofing and replication, 
as well as lightweight to be implemented on resource-constrained devices is crucial to the security and protection of such emerging systems. As a result, deep learning-enabled device fingerprinting has emerged as a promising technique 
for automatically identifying devices on the fly using physical-level features that can be captured automatically from received RF (radio frequency) signals and that are difficult, if not impossible, to spoof or replicate. 
In essence, these fingerprinting approaches rely on transceiver hardware impairments, inevitably inherited during manufacturing, that impair the transmitted RF signals in a way that provides transmitters with fingerprints that can uniquely separate them from one another~\cite{sankhe_oracle_2019}.

Although deep learning-based approaches have been proven efficient in identifying devices from captured RF signals, multiple studies (e.g.,~\cite{sankhe_oracle_2019,elmaghbub2021lora,restuccia_deepradioid_2019,hamdaoui2022deep}) indicate that these approaches suffer greatly when training and testing of the learning models are done over different wireless channels. In effect, recent experimental findings reveal that changes in the wireless channel conditions result in a drop of the device classification accuracy from $65$\% to $20$\% for LoRa data~\cite{elmaghbub2021lora} and from $85\%$ to $10\%$ for WiFi data~\cite{al-shawabka_exposing_2020}. 

There have been some efforts that attempted to overcome the impact of channel variations. 
For instance, Sankhe \textit{et al.} \cite{sankhe_oracle_2019} propose to modify the transmitting chains using software defined radios in a way that the demodulated symbols acquire unique characteristics that make the learning models more robust to channel changes. 
Elmaghbub \textit{et al.}~\cite{elmaghbub2021lora} leverage the out-of-band emissions in the band surrounding the original signal, caused by transceiver hardware impairments, to improve the insensitivity of device fingerprinting to channel changes. Using LoRa RF datasets collected using 25 IoT devices and an USRP B210 receiver~\cite{elmaghbub2021lora}, they show that the inherent out-of-band emissions increases the accuracy by $10$ to $20$\%.
Restuccia \textit{et al.} \cite{restuccia_deepradioid_2019}, on the other hand, show that finite input response (FIR) filters optimized to incur modification to the transmitted signals to account for current channel conditions can improve the accuracy from about $40\%$ to $60\%$. 
Although these proposed approaches overcome, to some extent, the channel variability issue, they require modification of the transmitted signals which results in increasing the bit error rates. In addition, most of these require changes to be made at the transmitters' side.

There has not been much done that exploits MIMO (multiple-input multiple-output) benefits to overcome channel variability in device fingerprinting. In~\cite{meneghello2022deepcsi}, the authors have recently proposed to exploit multi-user MIMO beamforming feedback matrices, computed by WiFi devices and sent back to the WiFi access point (AP), to extract AP-specific, interference-free fingerprints. This technique, however, relies on feedback to be sent by the receiver to the sender, works only for multi-user MIMO systems, and is intended for identifying resourceful WiFi APs.

In this paper, we expand the work in~\cite{9838976} and propose to leverage MIMO diversity to mitigate the impact of channel variability on deep learning-based device fingerprinting.
We emphasize that this paper is not concerned with the study of neural network models, nor with which deep learning models are best suited for RF fingerprinting. Instead, the contributions of our work lie in exploiting MIMO benefits to mitigate flat fading in Rayleigh channels, so as to improve the robustness of deep learning-based device fingerprinting to channel variations and automate the identification process of wireless devices in real-world scenarios. 
More specifically, we show that when compared to the conventional fingerprinting approach, using blind channel estimation enabled through combined MIMO and STBC (Space Time Block Coding) capabilities improves the device identification accuracy by up to $50\%$ when the learning models are trained and tested over the same Rayleigh flat fading channel conditions, and by up to $70\%$ when training and testing are done over different channels.

The rest of the paper is organized as follows. 
Section \ref{sec:mimo_background} provides a background on MIMO.
Section \ref{sec:proposed_approach} describes the device identification technique proposed to mitigate the impact of channel fading. 
Section~\ref{sec:dataset} presents the MIMO dataset used in this work.
Section~\ref{sec:cnn} describes the deep learning model and architecture used in our work.  
Section \ref{sec:Eval} presents the results and performance analysis. Section~\ref{sec:challenges} presents open challenges and Section \ref{sec:conc} concludes the paper.

\begin{figure*}
  \centering
  \includegraphics[width = 1\textwidth]{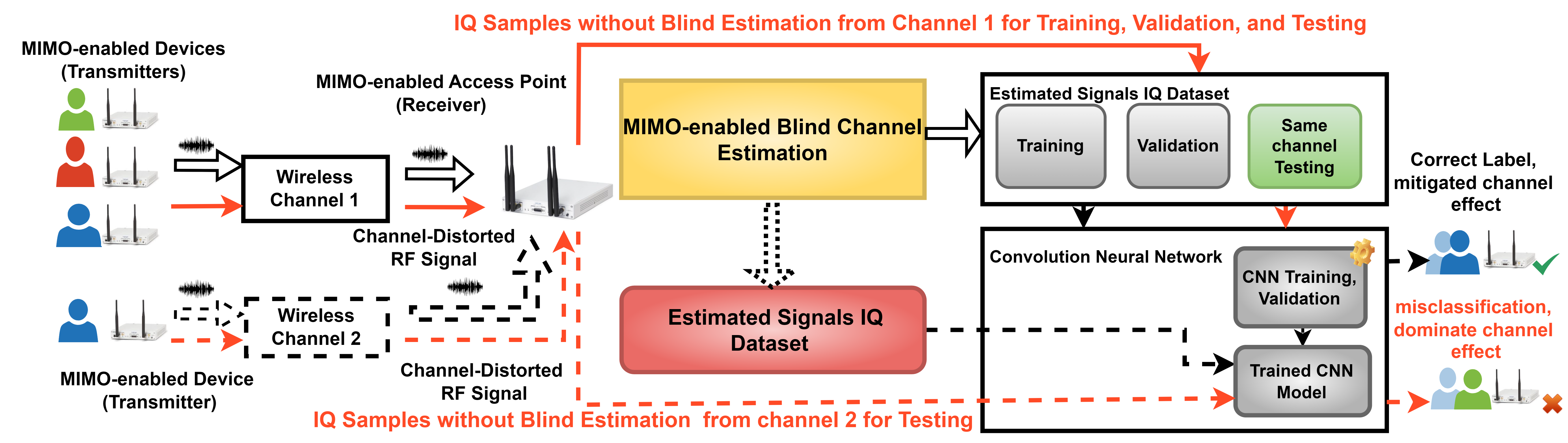}
    \caption{Overview of the proposed framework. Dashed arrows show the varying channel pipeline. Red arrows show the traditional SISO approach pipeline.}
    \label{fig:all}
\end{figure*}


\section{MIMO Diversity}
\label{sec:mimo_background}
MIMO is known to improve the Signal-to-Noise Ratio (SNR) through spatial diversity, and does so by combining the signals received on multiple uncorrelated antennas to overcome channel fading caused by multipath propagation~\cite{clerckx_mimo_2013}. This SNR improvement could be characterized by array gain capturing the increase in MIMO SNR relative to the single branch SNR, and by diversity gain capturing the increase in the slope of  the error rate as a function of the SNR~\cite{clerckx_mimo_2013}. Two types of diversity could be realized via MIMO. {\em Receive diversity}, which is realized through optimal combination of the received signals, and {\em transmit diversity}, which is enabled through MIMO's multiple transmit antenna capability, coupled with pre-processing or pre-coding capability offered, for example, through space-time block coding (STBC) techniques~\cite{clerckx_mimo_2013}. More specifically, STBC achieves transmit diversity by spreading information symbols in space using multiple transmitting antennas and in time through pre-coding~\cite{tarokh_space-time_1999-1, clerckx_mimo_2013}.

\section{Channel-Resilient Device Identification}
\label{sec:proposed_approach}
When a MIMO system transmits an STBC signal over a flat fading channel, each receiving antenna receives a signal combining all the signals transmitted by all the transmitting antennas. The contribution of each of the transmitted signals to the combined signal is a function of the channel condition observed between the corresponding transmitting and receiving antennas. As such, the process of estimating/recovering transmitted signals from received signal information, often referred to as blind source separation/blind channel estimation, boils down to determining the channel coefficients relating each of the transmitting and receiving antenna pair~\cite{ammar_blind_2002,ammar_blind_2007,shahbazpanahi_closed-form_2005}.

In this work, we leverage MIMO-enabled estimation capability to mitigate the distortions in the RF data caused by Rayleigh fading to improve the resiliency of device fingerprinting to channel variations and enable zero-touch identification of wireless devices in real-world scenarios.
More specifically, we consider two blind channel estimation approaches: MIMO-enabled full blind estimation (\MIMOFull) and MIMO-enabled partial blind estimation (\MIMOPartial). \MIMOFull~\cite{shahbazpanahi_closed-form_2005} blindly finds a closed-form estimation for the channel coefficient matrix using the orthogonal space-time block codes (OSTBC) properties and the second-order statistics of the received signals. \MIMOPartial~\cite{ammar_blind_2002, ammar_blind_2007}, on the other hand, partially estimates the channel matrix by showing that the unknown channel matrix lies in a subspace that can be determined from the received signal matrix given the STBC and the number of transmitting and receiving antennas. 
 
We use Tarokh STBC of rate $1/2$ with $3$ transmitting antennas and $3$ receiving antennas to transmit QPSK symbols to ensure that the channel matrix is identifiable up to a single complex ambiguity for \MIMOPartial~\cite{ammar_blind_2002, ammar_blind_2007}. A single complex ambiguity is expected to have a minor effect on the identification accuracy for \MIMOPartial. Once the channel coefficients are estimated using \MIMOFull~and \MIMOPartial, the transmitted signals are first recovered and then sampled and used for training the deep learning model to be used for device classification. 
Fig.~\ref{fig:all} depicts an overview of the proposed framework, showing its components and the different steps taken during the fingerprinting process, with and without blind estimation.

\section{RF Fingerprint Datasets}
\label{sec:dataset}
To evaluate the proposed approaches, we used MATLAB WLAN Toolbox to mimic WiFi-enabled devices. We changed the hardware impairments parameters of the devices, also using Matlab modeling, to emulate the impact of such impairments on the generated signals, and generated IEEE 802.11ac  waveforms for each of the simulated devices.
In our evaluation, we considered 20 devices, each having different impairment levels (IQ imbalance, Phase Noise, Frequency Offset, DC Offset). These impairments are purposely set slightly different across the different devices so as to mimic devices that have similar hardware, as this makes the classification task more challenging and hence can be used to validate the effectiveness of the proposed approaches vis-a-vis of their robustness to channel distortions. Refer to Table~\ref{Tb1} for specific details on the different impairment values used in our evaluation.

For each device, we first collected $5000$ frames, with each frame of size $160$. Then we split the real and imaginary parts of the signal and reshaped the frames as $2 \times 160$ vectors to be fed to the input layer of the CNN. The dataset was divided into $80\%$ for training, $10\%$ for validation and  $10\%$ for testing.

\begin{table*} [!t]
\centering 
\caption{\small Hardware impairments used to simulate 20 different devices. Devices are impaired with the low impairments set.} 
\resizebox{1\textwidth}{!}{\begin{tabular}{|l|l|l|l|l|l|l|l|l|} 
\hline 
Device & Phase Noise & Frequency Offset & IQ Gain Imbalance & IQ Phase Imbalance & AMAM & AMPM & Real DC Offset & Imaginary DC Offset\\
\hline 
DV1& -60 & 20 & 0.08 & 0.1 & [2.1587,1.1517] & [4.0033,9.104] & 0.1 & 0.15 \\
\hline
DV2& -60.15 & 20.01 & 0.1 & 0.09 & [2.1687,1.1617] & [4.1033,9.124]& 0.11 & 0.14 \\
\hline
DV3& -59.9& 20.2& 0.09 &0.09  & [2.1789,1.1317] & [4.0933,9.151] & 0.1 & 0.11 \\
\hline
DV4& -60.1 & 20  & 0.108 & 0.109 & [2.1987,1.1217] & [4.1033,9.194]  & 0.1 & 0.1 \\
\hline
DV5& -60  & 20.09& 0.1 & 0 & [2.1587,1.1717] & [4.093,9.094] & 0.089 & 0.1008\\
\hline
DV6& -59.95 & 20.1 & 0.12 & 0.15 & [2.1487,1.1117] & [4.1033,9.156] & 0.1 & 0.098  \\
\hline
DV7& -59.93&  20.11& 0.11 & 0.11 & [2.1897,1.1237] & [4.1133,9.135] & 0.111 & 0.1011 \\
\hline
DV8& -60.13 & 20.099 & 0.101 & 0.14 & [2.1387,1.1627] & [4.1533,9.096] & 0.12  & 0.099
 \\
\hline
DV9&-59.89  & 19.9& 0.099 & 0.08 & [2.1548,1.1917] & [4.09833,9.10056] & 0.09 & 0.0999 \\
\hline
DV10& -59.91 & 19.98  & 0.111 & 0.105 & [2.1777,1.09874] & [4.0987,9.123] & 0.101 & 0.10015 \\
\hline 
DV11& -60.16 & 20 & 0.1017 & 0.0899 & [2.1232,1.0999] & [4.0963,9.124] & 0.103 & 0.1006 \\
\hline
DV12& -60.09 & 19.8 & 0.1003 & 0.0913 & [2.1787,1.1236] & [4.1243,9.154]& 0.13 & 0.0996 \\
\hline
DV13& -59.99& 20.01& 0.0999 &0.0899  & [2.1987,1.1654] & [4.0935,9.0956] & 0.0999 & 0.1023 \\
\hline
DV14& -60.21 & 20.12  & 0.0992 & 0.0921 & [2.1569,1.1326] & [4.1253,9.199]  & 0.1002 & 0.1040 \\
\hline
DV15& -60.11  & 19.989& 0.1008 & 0.0941 & [2.1653,1.09876] & [4.1003,9.0988] & 0.0.1060 & 0.099\\
\hline
DV16& -60.123 & 20.09 & 0.1004 & 0.0841 & [2.0963,1.1207] & [4.1053,9.126] & 0.0890 & 0.0989  \\
\hline
DV17& -59.98&  20.123& 0.1006 & 0.0.0934 & [2.1456,1.1289] & [4.1111,9.105] & 0.1078 & 00.0988 \\
\hline
DV18& -59.88 & 20.18 & 0.1012 & 0.0863 & [2.1496,1.09627] & [4.1542,9.1366] & 0.1090  & 0.1010
 \\
\hline
DV19&-59.898  & 19.89& 0.1007 & 0.0895 & [2.1659,1.1097] & [4.0888,9.106] & 0.0987 & 0.1020 \\
\hline
DV20& -60.19 & 19.979  & 0.0995 & 0.0911 & [2.1967,1.09774] & [4.0999,9.0963] & 0.101 & 0.0959 \\
\hline
\end{tabular}}
\label{Tb1}
\end{table*}

\section{Deep Learning Model and Architecture}
\label{sec:cnn}
\label{sub:performance}
We used the convolution neural network (CNN) architecture used in~\cite{sankhe_oracle_2019} to assess the performance of the proposed techniques. It consists of two convolution layers and two fully connected layers. A $2 \times 160$ input is fed into the first convolution layer consisting of fifty $1 \times 7$ filters and producing $50$ features maps from the entire input. The second convolution layer has fifty $2 \times 7$ filters, with each filter being convoluted with the $50$-D volumes obtained from the first layer and serve to learn variations over the I and Q dimensions of the I/Q samples. The first, fully connected layer has $256$ nodes whose output are fed into the second fully connected layer. Both the convolution and fully connected layers use the ReLU activation function to add non-linearity. The last layer of the CNN is a softmax that yields the classification probabilities. The Adam optimizer is used to optimize the cross-entropy loss function evaluated at the classifier's output.

Note that although CNNs are widely recognized for their great RF fingerprinting performances and hence are also used in this work, other neural network models have also been used. That said, we want to emphasize that this work is not concerned with the design of new neural network models for RF fingerprinting, but instead focuses on improving the robustness of such existing models to channel variations. Hence, the proposed techniques are orthogonal to the neural network model used in the deep learning classification.

\section{Results and Performance Analysis}
\label{sec:Eval}
In this section, we assess the effectiveness of the studied MIMO-enabled approaches, \MIMOFull~and \MIMOPartial, for mitigating the impact of Rayleigh flat fading on the device identification and classification accuracy. 
The achieved performances are compared against the conventional SISO (single-input single-output) approach. 
We collect data from the 20 simulated WiFi devices, each enabled with a $3\times3$ MIMO to send QPSK symbols encoded via Tarokh STBC \cite{tarokh_space-time_1999-1} and impaired with a flat fading Rayleigh MIMO channel. At the receiving end, the reconstructed signals are sampled and used for training and testing the CNN-based approaches.

\subsection{Performance Metrics}\label{subsec:metrics}
Two metrics, {\em{Testing Accuracy}} and {\em{Testing Accuracy Gap}}, are used for the performance assessment. The latter metric represents the testing accuracy reduction (in percentage) resulting from using different channels for training and testing. That is, smaller gaps mean higher resiliency to channel variations.
We varied and studied the impact of the following parameters:
 
\begin{itemize}
    \item \textbf{Average Path Gain (APG)} of the Rayleigh fading channel used during training and testing.
    \item \textbf{Maximum Doppler Shift (MDS)} of the Rayleigh fading channel also used during training and testing. MDS is set to take on two values (0 Hz and 1 Hz) throughout.
    \item \textbf{Number of Devices (N)} to be identified, which is set to 10 and 20 to study its effect on the accuracy.
    \item \textbf{Impairments Set Type} capturing the impairment intensity via the standard deviation of the IQ (gain and phase) imbalances. Low standard deviation values mimic devices with similar/less distinguishable hardware, while high values mimic more distinguishable devices. 
\end{itemize}

We considered three impairments set types, Low, Moderate, and High, where the mean values of IQ-gain and IQ-phase imbalances are kept the same at $0.1$ and $0.09$, respectively, but the standard deviation values are varied as follows:
\begin{itemize}
        \item Low Intensity: 0.01 for gain and 0.02 for phase.
        \item Moderate Intensity: 0.055 for gain and 0.011 for phase.
        \item High Intensity: 0.1 for gain and 0.2 for phase.
\end{itemize}

\begin{figure}
	\centering
	\subfloat[Testing Accuracy (in \%)]{\includegraphics[width=1\columnwidth]{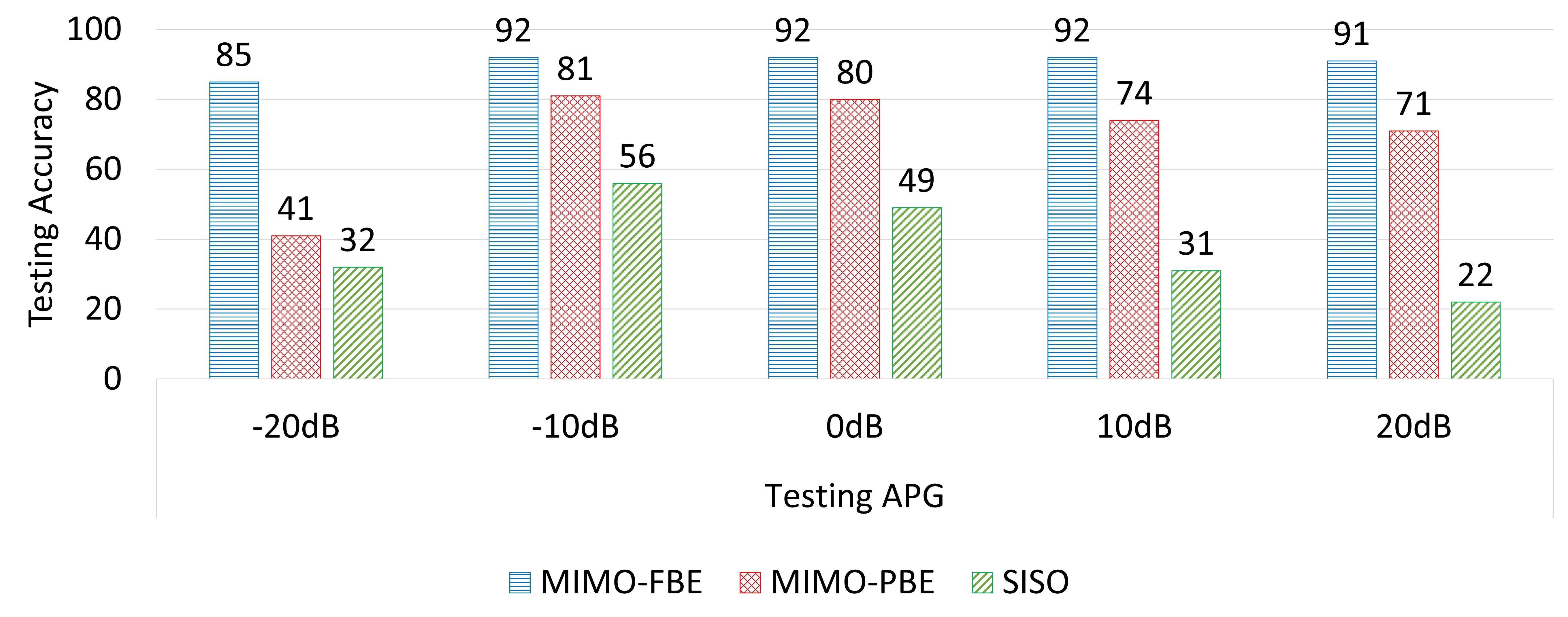}
    \label{subfig:acc}}
    
	\vspace{0.0001mm}
	\subfloat[Testing Accuracy Gap (in \%)]{
     \includegraphics[width=1\columnwidth]{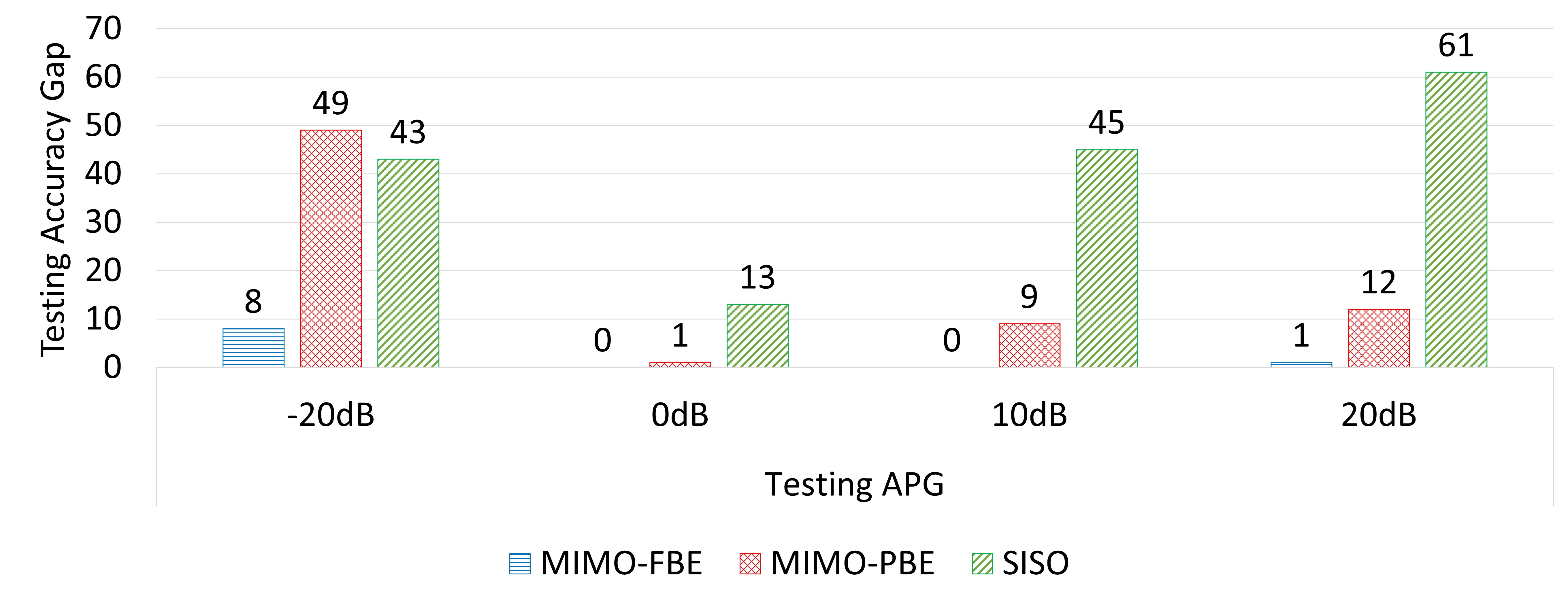} 
     \label{subfig:gap}}
    \caption{Impact of APG: Training MDS = 0Hz; Training APG = -10dB; Number of Devices N = 10; Impairments Set Type = low. \\
    Recall that: Testing Accuracy Gap (Testing APG = x dB) = [Testing Accuracy (Testing APG = -10 dB) -- Testing Accuracy (Testing APG = x dB)] / [Testing Accuracy (Testing APG = -10 dB)]}
    \label{fig:apg}
\end{figure}

\subsection{Impact of Channel Path Gain}
\label{sub:APG:Testing}
Fig.~\ref{fig:apg} shows the testing accuracy obtained over Rayleigh fading channels, where the CNN models used for all studied approaches are trained over a channel with APG = -10 dB and tested over a channel with different APG values, ranging from -20 to 20 dB. Recall that the Testing Accuracy Gap depicted in Fig.~\ref{subfig:gap} represents the normalized reduction in accuracy resulting from a testing APG being different from the training APG. The caption of Fig.~\ref{fig:apg} provides a precise definition.
These results allow us to make and confirm three key observations.

{\bf Sensitivity to channel variation.} First, observe the severe impact channel fading has on the obtained accuracy of the conventional SISO method. For example, as shown in Fig.~\ref{subfig:acc}, when training APG = -10 dB but testing APG = 20 dB, the accuracy of the SISO method drops from $56\%$ to $22\%$. This translates to a reduction gap in the accuracy of about $61\%$, as depicted in Fig.~\ref{subfig:gap}.
This observation demonstrates the severe sensitivity of the device identification process to changes in the channel conditions.
This is due to the distortion channel fading makes on the device fingerprints, making it too confusing for the learning models to identify the devices.

{\bf MIMO versus SISO.} Also, observe that the proposed MIMO-enabled approaches are much more resilient to channel fading than the conventional SISO approach. 
For instance, Fig.~\ref{subfig:acc} shows that when training APG = -10 dB but testing APG = 10 dB, while \MIMOFull~maintains an accuracy of $92\%$, the accuracy of SISO drops from $56\%$ to $31\%$, yielding an accuracy reduction gap of $45\%$, as indicated in Fig.~\ref{subfig:gap}. Also, observe that the proposed MIMO-enabled approaches show high resiliency even when the testing channel exhibits severe fading compared to the training channel. This can be seen from Fig.~\ref{subfig:gap}, which shows a negligible reduction gap in the accuracy achieved by \MIMOFull~even when the testing APG reaches 20 dB (with training APG still set to -10 dB). Note that \MIMOPartial~also maintains high resiliency to channel fading, though not as robust as \MIMOFull, but much more robust than SISO. In conclusion, these findings confirm that the use of blind estimation to find and compensate for the distortions imposed by channel fading improves the robustness of fingerprinting to channel condition variations.

Note that when the channel used for testing exhibit lesser APG (more severe fading) than the channel used for training, the resiliency of the MIMO-enabled approaches to channel variations degrades, especially for \MIMOPartial. This can be seen from Fig.~\ref{subfig:gap} which indicates that when testing APG = -20 dB (severe fading), the accuracy reduction gaps for \MIMOPartial~and SISO are $49\%$ and $43\%$, respectively. However, for \MIMOFull, the gap is only $8\%$, indicating a high channel resiliency of \MIMOFull~when compared to \MIMOPartial~and SISO. This is again because \MIMOFull~uses samples collected from the received signals that are reconstructed after estimating the channel coefficients.

{\bf Full versus partial blind estimation.} The final trend we observe is that \MIMOFull~outperforms \MIMOPartial. Note that when testing AGP = -10 dB, \MIMOFull~achieves a testing accuracy of $92\%$ compared to $81\%$ only for \MIMOPartial. Moreover, when both approaches are tested on channels with different testing APG values, \MIMOFull~is more resilient to channel variations. Note that when the training APG = -10 dB and the testing APG = 20 dB, \MIMOFull~achieves a testing accuracy of $91\%$ compared to only $71\%$ for \MIMOPartial. This decrease in the performance of \MIMOPartial~compared to \MIMOFull~is due to the remained (unresolved) ambiguity in the estimated coefficients of the channel in the case of \MIMOPartial.

\subsection{Impact of Maximum Doppler Shift (MDS)}
\begin{figure}
\includegraphics[width=1\columnwidth]{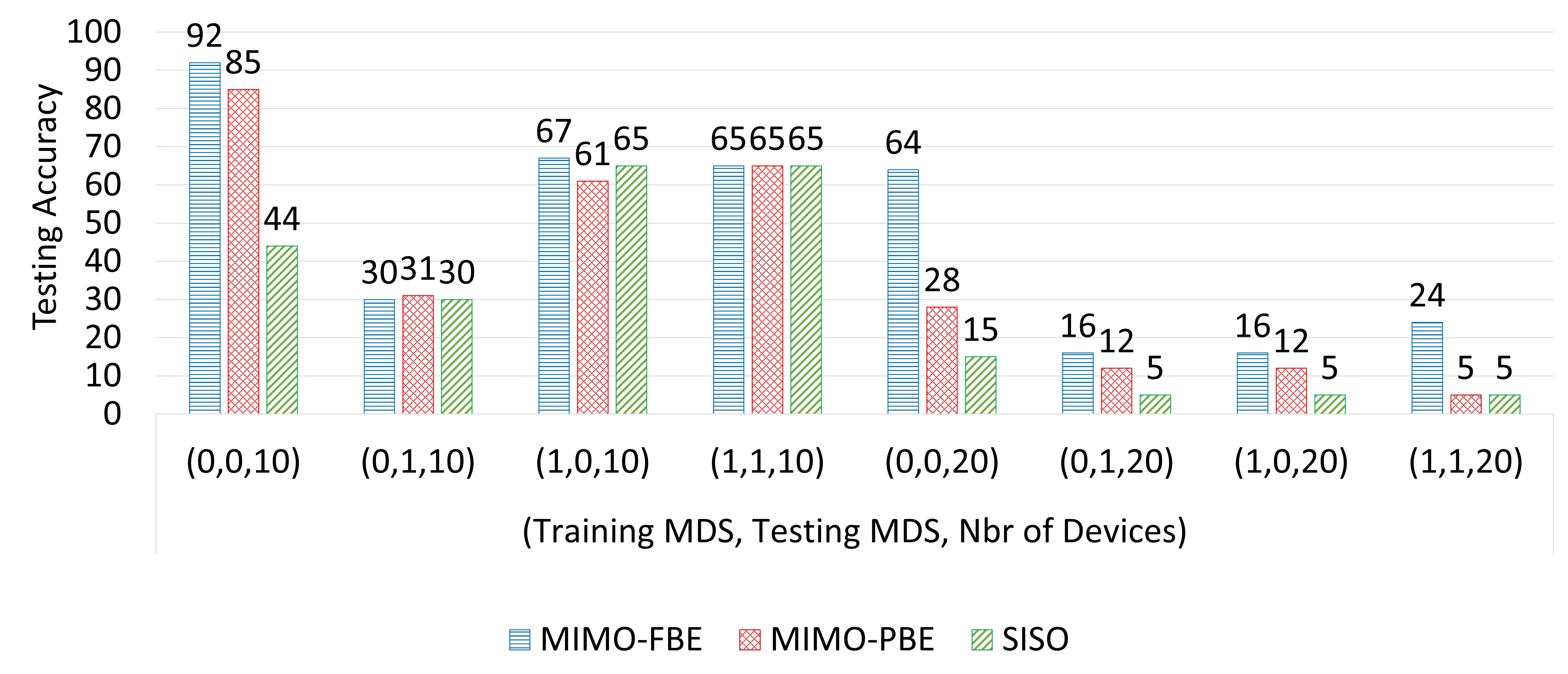}
\caption{Impact of MDS. Training APG = Testing APG = -20 dB.}
\label{fig:diff training doppler shifts-10}
\end{figure}

We now consider the dynamic scenario where the relative speed between the transmitter and the receiver, characterized by the channel maximum Doppler shift (MDS), changes. 
Fig.~\ref{fig:diff training doppler shifts-10} shows the testing accuracy achieved over Rayleigh channels when the CNN is trained and tested under two different MDS values (0 and 1 Hz). Each bin in the figure represents the testing accuracy obtained by all 3 studied approaches at a (Training MDS, Testing MDS, Number of Devices) combination. 
First, observe that for 10 devices when the training and testing MDS = 1 Hz, the testing accuracy for all 3 approaches is about $65\%$ compared to $92\%$, $85\%$, and $44\%$ for \MIMOFull, \MIMOPartial~and SISO when the training and testing MDS = 0 Hz. This result indicates the severe effect of the MDS on the RF fingerprinting accuracy even when the same MDS value is used for training and testing.
Second, observe that for 10 devices, when the training MDS = 0 Hz (static scenario) but the testing MDS = 1 Hz, the testing accuracy for \MIMOFull, \MIMOPartial, and SISO decreases from $92\%$, $85\%$, and $44\%$ to about $30\%$. The result shows that although the MIMO-enabled approaches can mitigate the channel fading effect, they fail to mitigate the effect of the relative speed between the transmitter and the receiver. 
Third, observe that increasing the number of devices to 20 while considering the dynamic scenario emphasizes the severe impact of MDS on RF fingerprinting. For instance, for 20 devices, when the training and testing MDS = 1 Hz, the testing accuracy for \MIMOFull~approach decreases to $24\%$ and both the \MIMOPartial~and the conventional SISO approaches are randomly guessing the devices' identities.

\subsection{Impact of the IQ Imbalance Intensity}
\begin{figure}
\includegraphics[width=1\columnwidth]{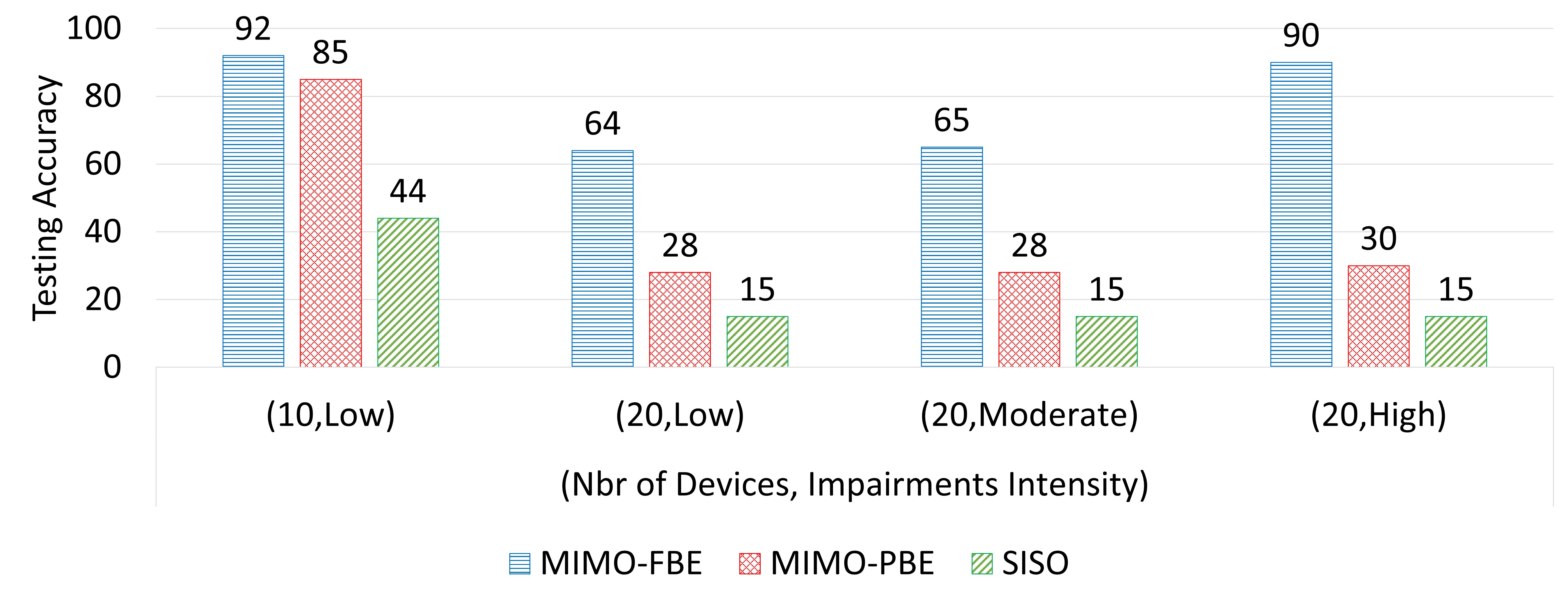}
\caption{Impact of IQ Imbalances. Training APG = Testing APG = -20 dB.}
\label{fig:impairementeffect}
\end{figure}

Fig.~\ref{fig:impairementeffect} shows the impact of the impairments intensity by considering the three IQ imbalance intensity levels shown in Section~\ref{subsec:metrics}: Low, Moderate and High. 
First, consider the case when the number of devices is 20. Observe that as the impairments intensity increases from Low to High, the accuracy achieved under \MIMOFull~increases from $64\%$ to $90\%$, whereas those achieved under \MIMOPartial~and SISO remain about $28\%$ and $15\%$, respectively, with no improvement over the Low impairments level. 
These findings show that \MIMOFull~outperforms \MIMOPartial~and SISO.
Recall that with the Low intensity level, devices tend to have similar impairment values, making them less distinguishable from one another, which explains why \MIMOFull~performs better under High intensity level. As for the superiority of \MIMOFull~to \MIMOPartial, this can be attributed to the remained ambiguities in the channel matrix when estimated by \MIMOPartial. 
That is, the reconstructed signals used for classification are more affected---and hence the impairments are more profoundly overshadowed---by fading. 

Fig.~\ref{fig:impairementeffect} also shows that when the number of devices increases from 10 to 20 while keeping the same intensity level (Low), \MIMOFull~accuracy decreases from $92\%$ to $64\%$, \MIMOPartial~accuracy decreases from $85\%$ to $28\%$, and SISO accuracy decreases from $44\%$ to $15\%$. The figure also shows that despite the degradation in the  accuracy, when the number of devices is doubled, both \MIMOFull~and \MIMOPartial~still outperform SISO, indicating the success of MIMO in mitigating the channel effect. Since fading results in signal distortion that overshadows the device fingerprints, a higher number of devices makes the fingerprints (feature vectors) closer to one another, thereby creating more confusion to the classifier and yielding lower identification accuracy.

\section{Unsolved Research Challenges}
\label{sec:challenges}
As shown in this paper, MIMO capabilities are proven to boost the device fingerprinting and identification resiliency against fading and channel condition variations. However, there remains key challenges and open research questions that need to be addressed. These include:
\begin{itemize}
    \item MIMO capabilities do not mitigate the impact of the relative speeds between the transmitters and the receiver. For instance, for crowded indoor environments, where the MDS values could exceed $30$ Hz at $3.6$ GHz, device classification and identification approaches still suffer from severe accuracy reduction~\cite{hanssens_measurement-based_2016}, and hence addressing the MDS effect on device fingerprinting remains unsolved and requires further investigation.
    
    \item Deep learning-enabled device fingerprinting still do not scale well, as the classification accuracy degrades as the number of devices increases. Developing approaches that leverage MIMO and other technologies to overcome scalability issues still needs research attention.
    
    \item Another untouched issue of great importance is the degradation of fingerprinting accuracy over time due, for instance, to hardware aging as well as variation of other conditions like temperature, which are expected to change over time, and so does their impact on fingerprinting accuracy. This has not been discussed in prior fingerprinting works, and it is thus crucial to determine how robust the classification models are to such changes.
   
    \item Device fingerprinting systems could be targeted by attackers who leverage inexpensive software defined radios to generate a handcrafted optimal attacker signal that interferes with the device RF signal to cause device misclassification. Attacks targeting machine learning models that are based on generating stealth perturbations to the neural network inputs are widely discussed in various contexts, like computer vision~\cite{su_one_2019}. However, the feasibility of such attacks in the RF fingerprinting context and how to secure against them remain open research problems.
\end{itemize}

\section{Conclusion}
\label{sec:conc}
In this paper, we proposed deep learning-based MIMO-enabled RF/device classification approaches that aim to overcome fading issues in Rayleigh channels. We showed that MIMO capabilities can help mitigate the wireless channel effect and improve the device fingerprinting and identification accuracy in Rayleigh flat fading channels. We also showed that although MIMO capabilities could increase the resiliency of these device fingerprinting approaches against channel gain variation, they could not overcome the effect of mobility between the transmitters and the receiver. The paper also highlights some of the key open issues that deep learning-based device fingerprinting still faces, including device scalability and machine learning model security.

\section*{Acknowledgment}
This work was supported in part by the US National Science Foundation under NSF awards 1923884 and 2003273.

\bibliographystyle{IEEEtran}
\bibliography{IEEEabrv,references}
\section*{Biography}
\begin{IEEEbiographynophoto}{Bechir Hamdaoui} [S’02-M’05-SM’12] (hamdaoui@eecs.oregonstate.edu) is a Professor in Computer Science and Engineering at Oregon State University. He received M.S. degrees in ECE (2002) and CS (2004), and Ph.D. degree in ECE (2005) from the University of Wisconsin-Madison. His research interests are in the general areas of intelligent networked systems, wireless and network security, and data communication networks. He won several awards, including the ISSIP 2020 Distinguished Recognition Award, the ICC 2017 Best Paper Award, the 2016 EECS Outstanding Research Award, and the 2009 NSF CAREER Award. He served as a Distinguished Lecturer for the IEEE Communication Society for 2016 and 2017, and as the Chair for the IEEE Communications Society’s Wireless Technical Committee (WTC) during the 2021-2022 term.
\end{IEEEbiographynophoto}
\begin{IEEEbiographynophoto}{Nora Basha}(bashano@oregonstate.edu) is currently a Ph.D. student at Oregon State University, USA, where she received her M.S.in ECE in 2021. Her research interests are wireless communication networks, data-enabled intelligent network access, and IoT and autonomous systems security. 
\end{IEEEbiographynophoto}
\begin{IEEEbiographynophoto} {Kathiravetpillai (Siva) Sivanesan}(kathiravetpillai.sivanesan@intel.com) is a Senior Research Scientist at Intel Labs. He is driving connected vehicles and Edge computing research for transportation and smart city applications. He earned his Ph.D. in wireless communications in 2004 and has authored more than 30 publications. He holds more than 70 patents, and has several patents pending.  
\end{IEEEbiographynophoto}
\end{document}